\documentclass[conference]{IEEEtran}
\IEEEoverridecommandlockouts
\usepackage{cite}
\usepackage{amsmath,amssymb,amsfonts}
\usepackage{algorithmic}
\usepackage{graphicx}
\usepackage{textcomp}
\usepackage{xcolor}
\usepackage{amsmath}
\usepackage{amssymb}
\usepackage{mathtools}
\usepackage{amsthm}
\usepackage{amsfonts}
\usepackage{stackengine}

\newtheorem{theorem}{Theorem}

\newtheorem{lemma}{Lemma}

\usepackage{float}
\def\BibTeX{{\rm B\kern-.05em{\sc i\kern-.025em b}\kern-.08em
    T\kern-.1667em\lower.7ex\hbox{E}\kern-.125emX}}
\begin{document}

\title{Downlink Pilots are Essential for Cell-Free Massive MIMO with Multi-Antenna Users
\thanks{This work was supported by the SUCCESS project funded by the Swedish Foundation for Strategic Research.}
}

\author{\IEEEauthorblockN{Eren Berk Kama}
\IEEEauthorblockA{\textit{Division of Communication Systems} \\
\textit{KTH Royal Institute of Technology}\\
Stockholm, Sweden \\
ebkama@kth.se}
\and
\IEEEauthorblockN{Junbeom Kim}
\IEEEauthorblockA{\textit{Department of Intell. Commun. Engineering} \\
\textit{Gyeongsang National University}\\
Tongyeong, South Korea \\
junbeom@gnu.ac.kr}
\and
\IEEEauthorblockN{Emil Björnson}
\IEEEauthorblockA{\textit{Division of Communication Systems} \\
\textit{KTH Royal Institute of Technology}\\
Stockholm, Sweden \\
emilbjo@kth.se}
}

\maketitle

\begin{abstract}
We consider a cell-free massive MIMO system with multiple antennas on the users and access points. In previous works, the downlink spectral efficiency (SE) has been evaluated using the hardening bound that requires no downlink pilots. This approach works well when having single-antenna users. In this paper, we show that much higher SEs can be achieved if downlink pilots are sent since the effective channel matrix does not harden when having multi-antenna users. We propose a pilot-based downlink estimation scheme and derive a new SE expression that utilizes zero-forcing combining. We show numerically how the number of users and user antennas affects the SE.
\end{abstract}


\section{Introduction}

Cell-free massive multiple-input multiple-output (MIMO) is a wireless communication paradigm that has attracted great interest due to the vision of delivering uniformly high spectral efficiency (SE) over the coverage area \cite{interdonato2019ubiquitous,chen2022survey,9650567}.
In cell-free massive MIMO, a large number of geographically distributed access points (APs) cooperate to serve the users on the same time-frequency resources through coherent joint transmission where interference is managed by MIMO methods \cite{demir2021foundations}.

The recent surveys \cite{9650567,chen2022survey} highlight a large amount of research on cell-free massive MIMO systems. The predominant assumption is single-antenna users, although user devices (e.g., phones and tablets) have been equipped with multiple antennas since 4G.
The few papers that consider multi-antenna users show SE improvements  (e.g., \cite{buzzi2017cell}) but assume perfect channel state information (CSI).
A detailed SE analysis with realistic imperfect CSI is missing, particularly for downlink operation.

The uplink SE with zero-forcing (ZF) combining and multi-antenna users was studied in \cite{mai2019uplink} under independent and identically distributed (i.i.d.) Rayleigh fading. This analysis was extended to consider the Weichselberger channel model in \cite{wang2022uplink}, where four operation regimes were compared. The uplink was further studied in \cite{zhang2019performance,sun2023uplink} with a focus on distortion caused by low-resolution hardware.
The downlink performance with ZF precoding analyzed in \cite{buzzi2019user} under i.i.d. Rayleigh fading, using the hardening bound where the user device lacks CSI. Moreover, the downlink performance with MR precoder and downlink channel estimation was studied in \cite{mai2020downlink}. This work was extended in \cite{zhou2021sum} to consider the downlink with low-resolution digital-to-analog converters (DACs).
Multi-antenna users have also been studied in the context of single-cell massive MIMO \cite{li2016massive}, but also using the hardening bound without CSI at the users.
 
In \cite{ngo2017no}, it was shown that no downlink pilots are needed in massive MIMO when considering single-antenna users. The receiver can perform decoding properly without CSI since the effective channel's phase is removed by precoding. The hardening bound gives performance close to perfect CSI also in cell-free massive MIMO \cite{demir2021foundations}. In cases with very limited channel hardening (e.g., keyhole channels), higher SE can be achieved by explicitly estimating the effective channel's amplitude from the received downlink signals \cite{ngo2017no}.
This might be the reason why the hardening bound is also used in \cite{buzzi2019user,zhou2021sum,li2016massive} for multi-antenna users. However, we will show that downlink pilots can give great improvements in these cases.

In this paper, we analyze the performance of a cell-free massive MIMO system with multiple antennas on the users and APs. We consider arbitrary precoding in the downlink and analyze the impact of different kinds of CSI. In particular, we derive an SE expression for the case when the user has CSI obtained through downlink channel estimation. As the effective channel is non-Gaussian, we use linear minimum mean-squared error (LMMSE) estimation to obtain the effective channel estimates. We further propose a way for the receiver to suppress interference using the estimates. The performance of this method is compared with the conventional hardening bound and the ideal case of having
perfect CSI at the receiver. 
We provide numerical results that show the importance of downlink pilots in this setup and demonstrate the impact of different numbers of antennas on the APs and users.

\section{System Model}
We consider a cell-free massive MIMO system with $L$ APs and $K$ users arbitrarily distributed in a large geographical area. Each AP has $N$ antennas and each user has $M$ antennas. We consider the standard block-fading TDD operation \cite{demir2021foundations}, where the channel between AP $l$ and user $k$ in an arbitrary coherence block is denoted by $\mathbf{H}_{lk} \in \mathbb{C}^{N \times M}$. The channel takes an independent realization in each block according to i.i.d. Rayleigh fading with variance $\beta_{lk}$; that is,  $\mathbf{H}_{lk}$ has i.i.d.   $\mathcal{N}_{\mathbb{C}}(0,\beta_{lk})$-entries. Similar to \cite{mai2019uplink,zhang2019performance, mai2020downlink, buzzi2019user,buzzi2017cell, zhou2021sum}, we consider this rich-scattering fading distribution to present the proposed downlink channel estimation procedures under analytically tractable conditions. However, the concepts can be generalized by inserting covariance matrices in the definitions of the channel matrices and making corresponding algebraic manipulations.

\subsection{Uplink Channel Estimation}

As customary in TDD operation, the APs estimate the channels in the uplink and use the estimates in both uplink and downlink.
All users send pilot sequences, and the APs estimate their channels to the users based on their received signals. Each user sends one $\tau_{\mathrm{p}}$ length orthonormal pilot sequence per transmit antenna. We will let $\tau_{\mathrm{c}}$  denote the coherence block (in number of symbols); thus, the pilot overhead is $\tau_{\mathrm{p}}/\tau_{\mathrm{c}}$.

We denote the pilot matrix used by user $k$ as $\boldsymbol{\Phi}_{k} \in \mathbb{C}^{\tau_{\mathrm{p}} \times M}$. We assume that more than one user uses the same pilot matrices, to limit the pilot overhead. We use the set $\mathcal{P}_{k}$ to denote the set of users that share pilots with user $k$. Pilot matrices are selected so that $\boldsymbol{\Phi}_{i}^{\rm{H}}\boldsymbol{\Phi}_{i'}= \mathbf{I}_{\tau_{\mathrm{p}}}$ if $i \in \mathcal{P}_{k}$ and $\mathbf{0}$ otherwise, and user $k$ transmits $\sqrt{\tau_{\mathrm{p}}} \boldsymbol{\Phi}_{k}$ to make the pilot energy proportional to the pilot length.  The received signal at AP $l$ is 
\begin{equation}
    \mathbf{Y}_{l}^{\text{Pilot}}=\sum_{i=1}^{K} \sqrt{ q_{i} \tau_{\mathrm{p}}} \mathbf{H}_{li} \boldsymbol{\boldsymbol{\Phi}}_{i}^{\rm{H}} + \mathbf{N}_{l},
\end{equation}
where $q_{i}$ is the transmit power used by user $i$ normalized by the noise power and $\mathbf{N}_{l}$ is the $M \times \tau_{\mathrm{p}}$ noise matrix at AP $l$ with i.i.d. $\mathcal{N}_{\mathbb{C}}(0,1)$-entries. After correlating the received signal with the pilot matrix of user $k$, we obtain
\begin{align} \nonumber
    \mathbf{Y}_{lk} &=\mathbf{Y}_{l}^{\text{Pilot}} \boldsymbol{\Phi}_{k} \\ \nonumber
    &= \sqrt{q_k \tau_{\mathrm{p}} } \mathbf{H}_{lk} + 
    \sum_{i=1}^{K} \sqrt{q_i  \tau_{\mathrm{p}}} \mathbf{H}_{li} \left( \boldsymbol{\Phi}_{i}^{\rm{H}}\boldsymbol{\Phi}_{k}\right) + 
    \mathbf{N}_{lk} \\
     &= \sqrt{q_k \tau_{\mathrm{p}}} \mathbf{H}_{lk} + \sum_{{i\in \mathcal{P}_{k}, i\neq k}} \sqrt{q_i \tau_{\mathrm{p}}} \mathbf{H}_{li} + \mathbf{N}_{lk},
\end{align}
where the second equality follows since the pilot  matrix $\boldsymbol{\Phi}_{i}$ is orthogonal to that of $\boldsymbol{\Phi}_{k}$ if $i \not \in \mathcal{P}_{k}$. We notice that the new noise matrix $\mathbf{N}_{lk}=\mathbf{N}_{l}\boldsymbol{\Phi}_{k}$ also has i.i.d. $\mathcal{N}_{\mathbb{C}}(0,1)$-entries; thus, every entry of $\mathbf{Y}_{lk}$ contains an observation of the corresponding Gaussian distributed entry of $\mathbf{H}_{lk}$ plus independent pilot interference and noise. The MMSE estimate of $\mathbf{H}_{lk}$ is \cite[Ch. 15]{kay1993fundamentals}

\begin{equation} \label{eq:Uplink_ChannelEstimation_last}
    \hat{\mathbf{H}}_{lk}=\frac{\beta_{lk}}{(q_k \tau_{\mathrm{p}} \sum_{{i\in \mathcal{P}_{k}}} \beta_{lk}^{2}  + 1 )}\mathbf{Y}_{lk} ,
\end{equation}
where we scale by the variance of each entry of $\mathbf{H}_{lk}$ and divide by the variance of each entry of $\mathbf{Y}_{lk}$.

\section{Downlink Data Transmission}

We focus on the downlink data transmission. All APs use the uplink channel estimates to precode the transmitted signals. We consider the centralized operation \cite{demir2021foundations}, where the APs cooperate in the transmission and share CSI. To simplify the notation, we stack all the AP channels to user $k$ in the $LN \times M$ matrix $\mathbf{H}_{k}= [ \mathbf{H}_{1k}^\intercal , \ldots  , \mathbf{H}_{Lk}^\intercal]^\intercal
$. Furthermore, we denote the $LN \times M$ precoding matrix of user $k$ as $\mathbf{W}_{k}= [ \mathbf{W}_{1k}^{\intercal} , \ldots  , \mathbf{W}_{Lk}^{\intercal}]^{\intercal}
$, where $\mathbf{W}_{lk} \in \mathbb{C}^{N \times M}$ is the part used by AP $l$. Hence, the transmitted data signal $\mathbf{x}_{k} \in \mathbb{C}^{N}$ meant for user $k$ is
\begin{equation}
    \mathbf{x}_{k}= \sum_{j=1}^{L} \mathbf{W}_{jk} \boldsymbol{\varsigma}_{k},
\end{equation}
where $\boldsymbol{\varsigma}_{k} \sim \mathcal{N}_{\mathbb{C}}(\mathbf{0},\mathbf{Q}_{k})$ is the data signal where $\mathbf{Q}_{k}$ is a diagonal matrix containing power control coefficients $q_{lk}$ between AP $l$ and user $k$. The power control coefficients are chosen so that $\mathbb{E} \left\{ \mathbf{x}_{k}^{\rm{H}} \mathbf{x}_{k} \right\} \leq \rho_d$, where $\rho_{d}$ is the transmit SNR constraint. The received signal at user $k$ is 
\begin{align} \label{downlink_transmitted_signal}
     \nonumber \mathbf{y}_{k} &=  \sum_{j=1}^{L}  \mathbf{H}_{jk}^{\rm{H}} \mathbf{W}_{jk} \boldsymbol{\varsigma}_{k} + \sum_{i=1, i \neq k}^{K} \sum_{j=1}^{L}  \mathbf{H}_{jk}^{\rm{H}} \mathbf{W}_{ji} \boldsymbol{\varsigma}_{i} +\mathbf{n}_{k} \\
     &=   \mathbf{H}_{k}^{\rm{H}} \mathbf{W}_{k} \boldsymbol{\varsigma}_{k} + \sum_{i=1, i \neq k}^{K}  \mathbf{H}_{k}^{\rm{H}} \mathbf{W}_{i} \boldsymbol{\varsigma}_{i} +\mathbf{n}_{k},
\end{align}
where $\mathbf{n}_{k} \sim \mathcal{N}_{\mathbb{C}}(\mathbf{0},\mathbf{I}_{M})$ is the normalized noise vector. The channel and precoding matrices appear in the form $\mathbf{B}_{ki}=\mathbf{H}_{k}^{\rm{H}} \mathbf{W}_{i}$ in \eqref{downlink_transmitted_signal}. We will use this notation for the effective channel after precoding in the remainder of this paper. For a given realization of the effective channels, created by the channel realizations and precoding scheme, the achievable SE varies depending on what CSI is available at the receiver. We will provide expressions for three different cases and then compare them numerically in Sec.~\ref{Numerical_Results}.

\subsection{Hardening Bound Without CSI at the Receiver} 

When the receiver lacks CSI regarding the instantaneous channel realizations, the SE can be lower bounded using the hardening bound \cite{demir2021foundations}. This method is conventionally used for single-antenna users, but the multi-antenna extension was utilized in \cite{buzzi2019user,zhou2021sum,li2016massive}. 
The main idea is that user $k$ knows the statistics of its effective channel $\mathbf{B}_{kk}$, such as the mean $\bar{\mathbf{B}}_{kk} = \mathbb{E}\left\{   \mathbf{B}_{kk} \right\}$. We can express the received signal in \eqref{downlink_transmitted_signal} as
\begin{equation} \label{eq:rewritten-hardening-bound}
     \mathbf{y}_{k} =  \bar{\mathbf{B}}_{kk} \boldsymbol{\varsigma}_{k} + \underbrace{(\mathbf{B}_{kk}-\bar{\mathbf{B}}_{kk}) \boldsymbol{\varsigma}_{k}+ \sum_{i=1, i \neq k}^{K} \mathbf{B}_{ki} \boldsymbol{\varsigma}_{i} +\mathbf{n}_{k}}_{\triangleq \mathbf{n}'_{k}},
\end{equation}
where $\mathbf{n}'_{k}$ denotes the sum of noise, the desired signal received over the unknown channel component, and inter-user interference. This term is uncorrelated with the first term $\bar{\mathbf{B}}_{kk} \boldsymbol{\varsigma}_{k}$ in \eqref{eq:rewritten-hardening-bound}, but it is spatially colored with the covariance matrix 
\begin{equation}
     \mathbf{\Xi}_{k} =  \mathbb{E}\left\{   \mathbf{n}'_{k} \mathbf{n}'^{\rm{H}}_{k} \right\}. 
\end{equation}

\begin{lemma} \label{first_lemma}
In the absence of CSI, an achievable SE at user $k$ is
\begin{equation} \label{first_theorem}
    \mathrm{SE}_{k}^\mathrm{noCSI}=\left(1 - \frac{\tau_{\mathrm{p}}}{\tau_{\mathrm{c}}}\right) \log_{2} \left|\mathbf{I}_{M} + \mathbf{Q}_{k}\bar{\mathbf{B}}_{kk}^{\rm{H}} \mathbf{\Xi}_{k}^{-1} \bar{\mathbf{B}}_{kk}\right|,
\end{equation}
where $| \cdot |$ denotes the determinant.
\end{lemma}

The proof of this lemma builds on interpreting \eqref{eq:rewritten-hardening-bound} as a deterministic point-to-point MIMO system with the channel matrix $\bar{\mathbf{B}}_{kk}$ and uncorrelated additive noise with the covariance matrix $\mathbf{\Xi}_{k}$. The SE expression is then obtained as a lower bound on capacity by utilizing the worst-case uncorrelated additive noise theorem \cite{Hassibi2003a}.
The single-antenna version of Lemma~\ref{first_lemma} is called the hardening bound \cite{demir2021foundations} since the effective channel realization $\mathbf{B}_{kk}$ is close to its mean  $\bar{\mathbf{B}}_{kk}$ when $M=1$ and $N$ is large. However, as the effective channel is a matrix instead of a scalar when $M>1$, the same kind of channel hardening is not achieved since each column of the precoding matrix can only achieve hardening with respect to one row in the channel matrix. The resulting performance loss will be demonstrated later in this paper.

To showcase the lack of channel hardening, we will provide a simple example with MR precoding based on perfect CSI and a single user. The $m$th row of effective channel between AP $l$ and user $k$ is 
\begin{align} \label{eq:hardening-mth-row}
    \mathbf{h}_{lk,m}^{\rm{H}}\mathbf{H}_{lk} &=  [ \mathbf{h}_{lk,m}^{\rm{H}}\mathbf{h}_{lk,1} , \ldots \mathbf{h}_{lk,m}^{\rm{H}} \mathbf{h}_{lk,m} , \ldots \mathbf{h}_{lk,m}^{\rm{H}} \mathbf{h}_{lk,M}],
\end{align}
where $\mathbf{h}_{lk,m}$ is the $m$th column of $\mathbf{H}_{lk}$. The $m$th term in \eqref{eq:hardening-mth-row} is
$\mathbf{h}_{lk,m}^{\rm{H}} \mathbf{h}_{lk,m}=\|\mathbf{h}_{lk,m} \|^{2} $ and has the asymptotic behavior 
\begin{align} \label{eq:channel_hardening}
      \frac{\|\mathbf{h}_{lk,m} \|^{2} }{\mathbb{E}\left\{\|\mathbf{h}_{lk,m} \|^{2} \right\}} \rightarrow 1,
\end{align}
in the mean square sense as $N \rightarrow \infty$ \cite[Ch. 2]{bjornson2017massive}, which is the classical channel hardening property. With similar reasoning, all the other elements in \eqref{eq:hardening-mth-row} will not harden because they are products of independent vectors. Hence, using so-called favorable propagation arguments \cite[Ch. 2]{bjornson2017massive}, it holds that
\begin{align}
      \frac{1}{\mathbb{E}\left\{\|\mathbf{h}_{lk,m} \|^{2} \right\}}[ \mathbf{h}_{lk,m}^{\rm{H}}\mathbf{h}_{lk,1} , \ldots \mathbf{h}_{lk,m}^{\rm{H}} \mathbf{h}_{lk,m} , \ldots \mathbf{h}_{lk,m}^{\rm{H}} \mathbf{h}_{lk,M}] \rightarrow \nonumber\\
     [ 0 , \ldots 1  , \ldots 0]
\end{align}
as $N \rightarrow \infty$.
This relation repeats on each row of the effective channel. Noting that the effective channel $\mathbf{B}_{kk}$ is the sum of effective channels between APs and user $k$, we conclude that only its diagonal elements achieve channel hardening. There is only one entry when $M=1$, so channel hardening holds for it, while most entries do not harden when $M>1$. Therefore, the hardening bound is loose when $M>1$, which calls for acquiring CSI at the receiving user.

\subsection{Capacity Bound with Perfect CSI at the Receiver}

Next, we consider the ideal case with perfect knowledge of the effective channels at the receiver. The CSI is obtained in a genie-aided way, which makes this case an upper bound that we can compare with other practical methods. We write the downlink received signal in \eqref{downlink_transmitted_signal} as
\begin{equation} 
\begin{split}
     \mathbf{y}_{k} =   \mathbf{B}_{kk} \boldsymbol{\varsigma}_{k} + \mathbf{n}''_{k},
\end{split}
\end{equation}
where $\mathbf{n}''_{k} = \sum_{i=1, i \neq k}^{K}  \mathbf{H}_{k}^{\rm{H}} \mathbf{W}_{i} \boldsymbol{\varsigma}_{i} +\mathbf{n}_{k}$ consists of the independent noise and interference terms.

\begin{lemma} \label{second_lemma}
With perfect CSI, an achievable SE at user $k$ is 
\begin{equation} \label{Perfect_CSI_SE}
    \mathrm{SE}_{k}^\mathrm{fullCSI} = \left(1 - \frac{\tau_{\mathrm{p}}}{\tau_{\mathrm{c}}}\right) \mathbb{E}\left\{  \log_{2} \left|  \mathbf{I}_{M} +  \mathbf{Q}_{k}\mathbf{B}_{kk}^{\rm{H}}\Tilde{\mathbf{\Xi}}_{k}^{-1} \mathbf{B}_{kk}  \right| \right\},
\end{equation}
where the covariance matrix $\Tilde{\mathbf{\Xi}}_{k}$ of $\mathbf{n}''_{k}$ is
\begin{equation}
     \Tilde{\mathbf{\Xi}}_{k} =  \sum_{i=1, i \neq k}^{K}  \mathbf{B}_{ki} \mathbf{B}_{ki}^{\rm{H}} +\mathbf{I}_{M}.
\end{equation}
\end{lemma}

This result is proved by treating the term $\mathbf{n}''_{k}$ as worst-case Gaussian noise, whitening the noise, and then stating the ergodic SE. We will later show by simulations that there is a large gap between the SE with perfect CSI in Lemma~\ref{second_lemma} and without CSI at the receiver in Lemma~\ref{first_lemma}. Therefore, we will analyze downlink channel estimation as a means to improve the SE.

\subsection{Estimation of the Effective Downlink Channel}

A viable way to provide the receiver with CSI is to send pilots using the downlink precoding from which user~$k$ can estimate the effective channel $\mathbf{B}_{kk}$. Since the effective channel takes random non-Gaussian realizations from a stationary distribution, a suitable estimation method is LMMSE estimation \cite[Ch. 12]{kay1993fundamentals}. To enable the estimation, the APs jointly send orthonormal pilot sequences using the selected precoding. The pilot matrix assigned to user $k$ is $\boldsymbol{\tilde{\Phi}}_{k} \in \mathbb{C}^{\tau_{\mathrm{p}} \times M}$. Again, $\mathcal{P}_{k}$ is the set of users that use the same pilot matrix as user $k$. The pilot matrices are orthonormal, that is, $\boldsymbol{\tilde{\Phi}}_{i}^{\rm{H}} \boldsymbol{\tilde{\Phi}}_{i'} = \mathbf{I}_{\tau_{\mathrm{p}}}$ if $i \in \mathcal{P}_{k}$ and $\boldsymbol{\tilde{\Phi}}_{i}^{\rm{H}} \boldsymbol{\tilde{\Phi}}_{i'} = \mathbf{0}$ otherwise. Note that we use the same pilot length $\tau_{\mathrm{p}}$ as in the uplink, but in principle one could use a different length. By sending these pilots over the channel in \eqref{downlink_transmitted_signal}, the received signal at user $k$ becomes

\begin{equation}
    \tilde{\mathbf{Y}}_{k}^{\text{Pilot}}=\sum_{i=1}^{K} \sqrt{q_i \tau_{\mathrm{p}}} \mathbf{B}_{ki} \boldsymbol{\tilde{\Phi}}_{i}^{\rm{H}} + \mathbf{N}_{k}.
\end{equation}
 By correlating with the dedicated pilot for user $k$, we obtain
\begin{align}   
    \tilde{\mathbf{Y}}_{k} &=\tilde{\mathbf{Y}}_{k}^{\text{Pilot}} \boldsymbol{\tilde{\Phi}}_{k}=\sum_{i=1}^{K} \sqrt{q_i \tau_{\mathrm{p}}} \mathbf{B}_{ki} \boldsymbol{\tilde{\Phi}}_{i}^{\rm{H}}  \boldsymbol{\tilde{\Phi}}_{k} + \mathbf{N}_{k} \nonumber \\ \label{downlink_channel_estimate0}
     &= \sqrt{q_k \tau_{\mathrm{p}}} \mathbf{B}_{kk} + \sum_{{i\in \mathcal{P}_{k}}, i\neq k} \sqrt{q_i \tau_{\mathrm{p}}} \mathbf{B}_{ki} + \mathbf{N}_{k}.
\end{align}
The entries of these matrices are statistically correlated. To describe the correlation using covariance matrices, we first need to 
vectorize the observation equation in \eqref{downlink_channel_estimate0}:
\begin{equation} \label{downlink_channel_estimate}
\begin{split}
    \mathrm{vec}(\tilde{\mathbf{Y}}_{k}) &= \sqrt{q_k \tau_{\mathrm{p}}} \mathrm{vec}(\mathbf{B}_{kk}) + \!\!\!\! \sum_{{i\in \mathcal{P}_{k}}, i \neq k} \!\!\!\sqrt{q_i \tau_{\mathrm{p}}} \mathrm{vec}(\mathbf{B}_{ki}) + \mathbf{n}_{k} ,
\end{split}
\end{equation}
where $\mathbf{n}_{k} = \mathrm{vec}(\mathbf{N}_{k})$.
We will let $\mathbf{b}_{kk} = \mathrm{vec}(\mathbf{B}_{kk})$ and $\mathbf{\tilde{y}}_{k} = \mathrm{vec}(\tilde{\mathbf{Y}}_{k})$ denote the vectorized effective channel and received signals, respectively. The LMMSE estimate of $\mathbf{b}_{kk}$ is 
\begin{equation} \label{eq:LMMSE_estimator}
    \hat{\mathbf{b}}_{kk} =   \mathbb{E}\left\{ \mathbf{b}_{kk} \right\} +  \mathbf{C}_{\mathbf{b}_{kk} \mathbf{\tilde{y}}_{k}}\mathbf{C}_{\tilde{\mathbf{y}}_{k}}^{-1}    (\mathbf{\tilde{y}}_{k} - \mathbb{E}\left\{\mathbf{\tilde{y}}_{k} \right\} ),
\end{equation}
where the covariance matrices $\mathbf{C}_{\mathbf{b}_{kk} \mathbf{\tilde{y}}_{k}},\mathbf{C}_{\mathbf{b}_{kk}}, \mathbf{C}_{\mathbf{n}_{lk}}$ are cross-covariance of effective channel and observation, the covariance of effective channel and the covariance of noise respectively. The estimation error covariance of this LMMSE estimator is
\begin{equation}
\mathbf{C}_{\tilde{\mathbf{b}}_{kk}} = \mathbf{C}_{\mathbf{b}_{kk}} - \mathbf{C}_{\mathbf{b}_{kk} \mathbf{\tilde{y}}_{k}} \mathbf{C}_{\mathbf{\tilde{y}}_{k}}^{-1}
\mathbf{C}_{\mathbf{b}_{kk} \mathbf{\tilde{y}}_{k}}
\end{equation}
and the total MSE is 
$\text{MSE} = \text{tr}(\mathbf{C}_{\tilde{\mathbf{b}}_{kk}})$.

\subsection{Capacity Bound with Estimate Channel at the Receiver}

We will now derive an achievable SE expression for the case when the receiver uses the proposed LMMSE estimator to acquire the effective channel matrix. Since the estimate $\hat{\mathbf{b}}_{kk}$ and estimation error $\mathbf{b}_{kk}-\hat{\mathbf{b}}_{kk}$ are non-Gaussian, they are uncorrelated but statistically dependent which rules out the use of classical capacity bounds. To address this issue, we propose to apply a ZF combining receiver that makes the channel nearly deterministic and thereby mitigates the dependencies.
Applying a general receive combining matrix $\mathbf{U}_{k} \in \mathbb{C}^{M \times M}$ to \eqref{downlink_transmitted_signal}, we obtain
\begin{equation} \label{eq:observation_with_ZF}
      \tilde{\mathbf{y}}_{k} =  \mathbf{U}_{k}^{\rm{H}} \mathbf{y}_{k} =  \mathbf{U}_{k}^{\rm{H}}  \mathbf{B}_{kk} \boldsymbol{\varsigma}_{k} + \mathbf{U}_{k}^{\rm{H}} \sum_{i=1, i \neq k}^{K} \mathbf{B}_{ki} \boldsymbol{\varsigma}_{i} + \mathbf{U}_{k}^{\rm{H}} \mathbf{n}_{k}.
\end{equation}
If we let $\hat{\mathbf{B}}_{kk}$ denote the LMMSE estimate as a matrix, ZF combining is given by $\mathbf{U}_{k}^{\rm{H}} = (\hat{\mathbf{B}}_{kk}^{\rm{H}} \hat{\mathbf{B}}_{kk})^{-1} \hat{\mathbf{B}}_{kk}^{\rm{H}}$. By expressing the unknown part of the channel as $\tilde{\mathbf{B}}_{kk} = \mathbf{B}_{kk} - \hat{\mathbf{B}}_{kk}$, \eqref{eq:observation_with_ZF} can be expressed as
\begin{equation} \label{eq:observation_equation_afterZFcombiner}
\begin{split}
      \tilde{\mathbf{y}}_{k} &=  (\hat{\mathbf{B}}_{kk}^{\rm{H}} \hat{\mathbf{B}}_{kk})^{-1} \hat{\mathbf{B}}_{kk}^{\rm{H}}\hat{\mathbf{B}}_{kk} \boldsymbol{\varsigma}_{k} + \\ 
      &  \quad (\hat{\mathbf{B}}_{kk}^{\rm{H}} \hat{\mathbf{B}}_{kk})^{-1} \hat{\mathbf{B}}_{kk}^{\rm{H}} \Big(  \tilde{\mathbf{B}}_{kk} \boldsymbol{\varsigma}_{k} + \sum_{i=1, i \neq k}^{K} \mathbf{B}_{ki} \boldsymbol{\varsigma}_{i} +  \mathbf{n}_{k} \Big) \\
      &=   \boldsymbol{\varsigma}_{k} + \underbrace{(\hat{\mathbf{B}}_{kk}^{\rm{H}} \hat{\mathbf{B}}_{kk})^{-1} \hat{\mathbf{B}}_{kk}^{\rm{H}} \Big(  \tilde{\mathbf{B}}_{kk} \boldsymbol{\varsigma}_{k} + \sum_{i=1, i \neq k}^{K} \mathbf{B}_{ki} \boldsymbol{\varsigma}_{i} +  \mathbf{n}_{k} \Big)}_{\triangleq \mathbf{n}'_{k}}.
\end{split}
\end{equation}
We notice that this is the transmitted signal $\boldsymbol{\varsigma}_{k}$ plus the uncorrelated interference and noise term $\mathbf{n}'_{k}$. We can now establish the following main result.

\begin{theorem} \label{th:mainresult}
When downlink pilots, the LMMSE estimator, and ZF combining are utilized, an achievable SE at user $k$ is
\begin{equation} \label{theorem_3}
     \mathrm{SE}_{k}^\mathrm{pilots} = \left(1 - \frac{2\tau_{\mathrm{p}}}{\tau_{\mathrm{c}}}\right) \log_{2} \left| \mathbf{I}_{M} + \mathbf{Q}_{k}\mathbf{C}_{\mathbf{n}'_{k}}^{-1}  \right|, 
\end{equation}
where $\mathbf{C}_{\mathbf{n}'_{k}}$ is the covariance of the term defined in \eqref{eq:observation_equation_afterZFcombiner}.
\end{theorem}
\begin{IEEEproof}
    The proof is given in Appendix A.
\end{IEEEproof}

When comparing this new SE expression to Lemma \ref{first_lemma}, we notice that the new pre-log factor is smaller since it compensates for the fact that pilots are transmitted in both uplink and downlink. The performance benefit comes from the matrix expression inside the determinant, where Theorem~\ref{th:mainresult} will give a larger SE since we use the downlink channel estimate to equalize the received signal as in \eqref{eq:observation_equation_afterZFcombiner}. We will demonstrate the performance improvement numerically in the next section.

\section{Numerical Results} \label{Numerical_Results}

In this section, we provide numerical results where we will compare the downlink SEs achieved with different CSI at the users and different numbers of antennas and APs. We use the same simulation setup as in \cite{demir2021foundations}.
We assume that the AP and user locations are uniformly distributed in a $1 \times 1 \; \text{km}^2$ area.  We use mutually orthogonal pilot sequences with pilot sharing among users. We use $K'$ to denote the number of orthogonal pilot matrices. $K/K'$ pilot-sharing users are chosen uniformly at random. The pilot length is $\tau_{\mathrm{p}} = K' M$ for uplink and for downlink channel estimation.  Equal power allocation is used between the users and their $M$ data streams. For the simulations, MMSE precoder is used in data transmission. Precoder for user $k$ is given as
\begin{equation}
    \mathbf{W}_{k}= \left[ \Big(\sum_{i=1}^{K}  \hat{\mathbf{H}}_{i}\hat{\mathbf{H}}_{i}^{H}  + \text{cov}\big(\tilde{\mathbf{H}}_{i}\big)\Big) +  \mathbf{I}_{LN}  \right] ^{-1} \hat{\mathbf{H}}_{k},
\end{equation}
where $\text{cov}\big(\tilde{\mathbf{H}}_{i}\big)$ is the covariance of the uplink channel estimation error for user $i$. The precoders are normalized so that per-AP power constraints are satisfied. 

\begin{figure}[t!]
\centering
  \includegraphics[width=\linewidth]{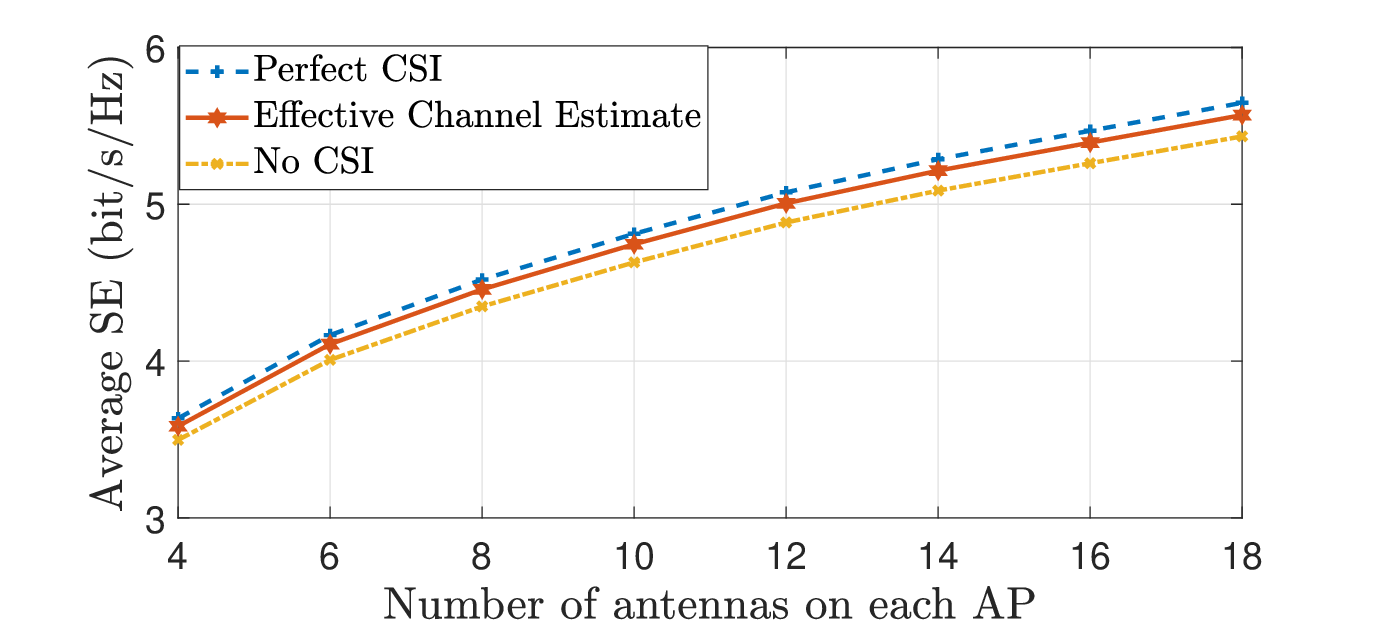} \vspace{-6mm}
  \caption{Achievable per user SE as a function of the number of AP antennas with $M=1$.} \vspace{-3mm}
  \label{fig:figureA}
\end{figure}

\begin{figure}[t!]
\centering
  \includegraphics[width=\linewidth]{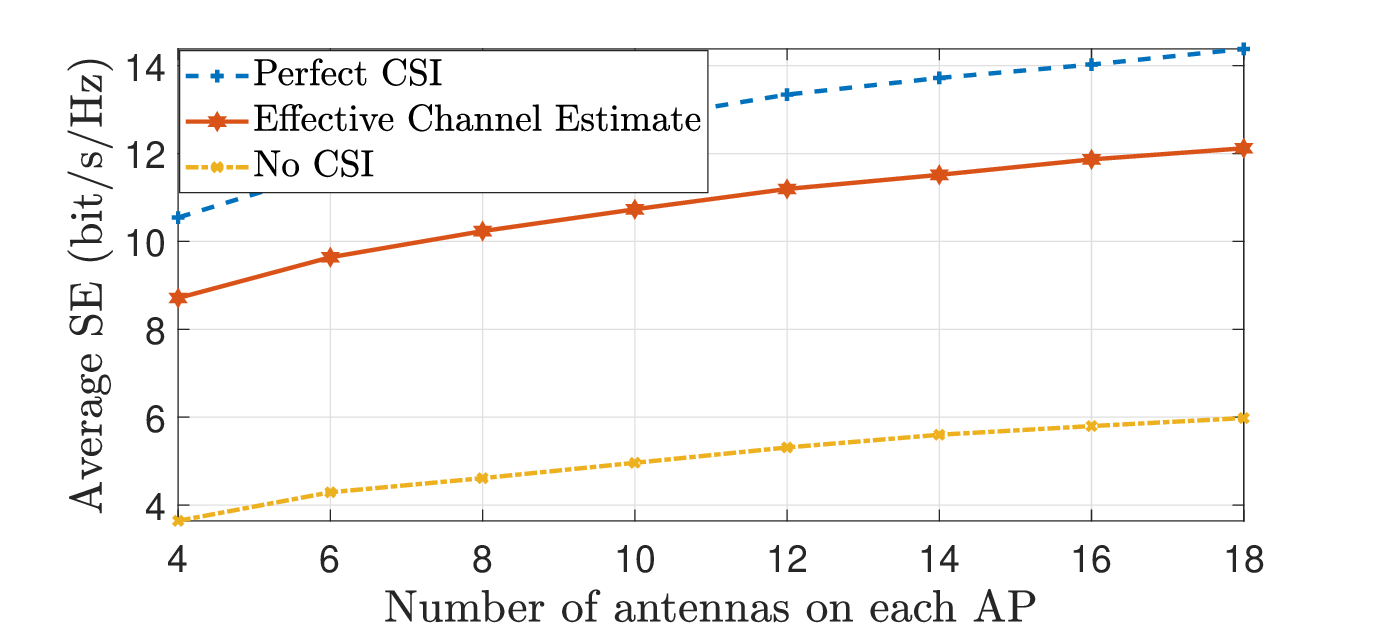} \vspace{-6mm}
  \caption{Achievable per user SE as a function of the number of AP antennas with $M=2$.} \vspace{-3mm}
  \label{fig:figure1}
\end{figure}

\begin{figure}[t!]
\centering
  \includegraphics[width=\linewidth]{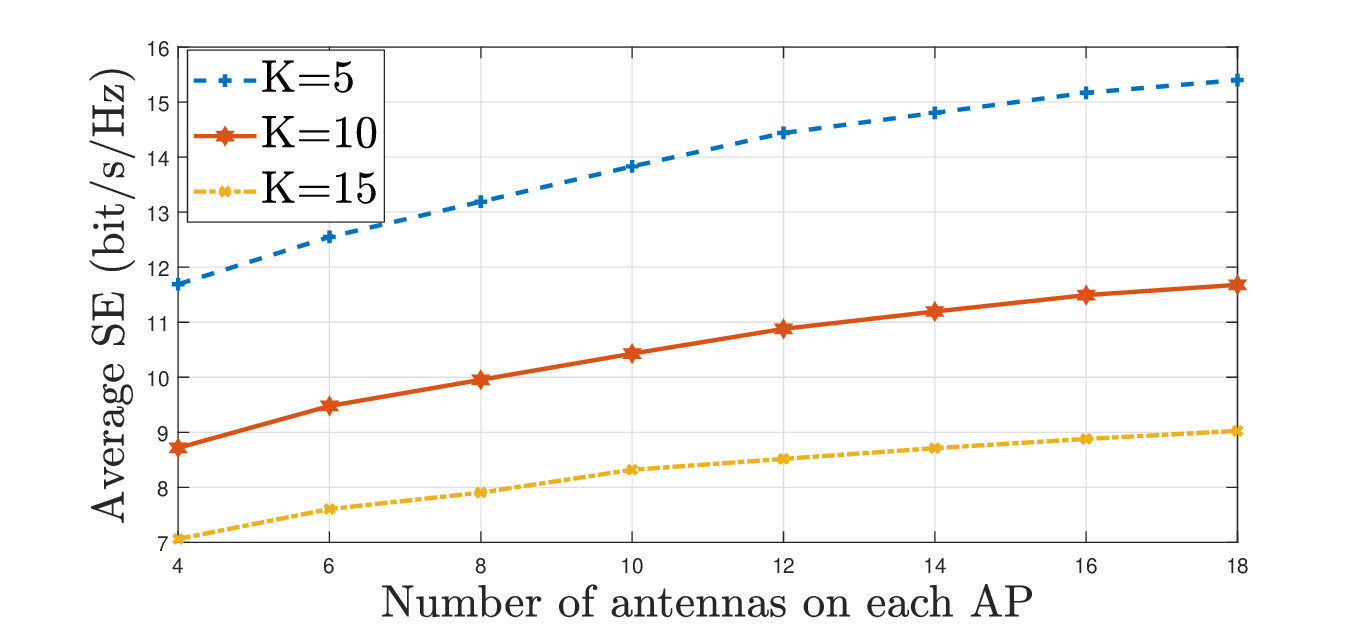} \vspace{-6mm}
  \caption{Achievable per user SE as a function of the number of users and user antennas.}
  \label{fig:figure3}
\end{figure}

Fig.~\ref{fig:figureA} and Fig.~\ref{fig:figure1} show the SE achieved in the three CSI cases: no CSI, perfect CSI, and downlink pilots to obtain effective channel estimates.
Both figures consider $L=10$ APs and $K=5$ users. These figures consider $M=1$ and $M=2$ antennas, respectively.
Moreover, Fig.~\ref{fig:figure3} shows the SE per user with downlink pilots when having $L=20$ APs and a varying number of users with $M=2$ antennas on users.

In Fig.~\ref{fig:figureA}, we see that with single-antenna users, there isn't much need for downlink pilots since the three curves are closely spaced. This result is stated in \cite{ngo2017no} for cellular massive MIMO and shown numerically in \cite{demir2021foundations} for cell-free massive MIMO.  On the other hand, in Fig.~\ref{fig:figure1}, we see that the gap between the curves is large when the user has multiple antennas. In particular, the proposed scheme with downlink pilots provides much greater SEs than the conventional approach without CSI at the receiver.
We also observe that the proposed approach is close to the genie-aided perfect CSI case. We observe that as the number of antennas on APs increases, the per-user SE increases thanks to the array gain. 

Fig.~\ref{fig:figure3} shows that, as more users are added, the performance per user drops.  The reason is the additional interference between the users. As the MMSE precoder suppresses interference, the performance is not severely affected. For example, the sum rate with $N=4$ increases by around $2.6 \times$ and $5 \times$ when going from $K=5$ to $K=10$ and $K=15$, respectively.

\section{Conclusions}

We have taken a new look at the downlink of cell-free massive MIMO with multi-antenna users. While good performance can be achieved without downlink pilots when having single-antenna users, this is not the case when having multi-antenna users. We proposed a new pilot-based downlink LMMSE channel estimation scheme and derived a novel SE expression. Through simulations, we observed that using downlink channel estimation significantly improves the SE, and the proposed method performs close to the perfect CSI case. We have also showcased the effect of varying numbers of users, APs, and user and AP antennas. 

\appendix

\section{Appendix}

\subsection{Proof of Theorem \ref{th:mainresult}}
The channel capacity with knowledge of $\bar{\mathbf{B}}_{kk}$ is defined as
\begin{equation} \label{capacity_definition}
    C = \max_{p(\mathbf{x}_{k})} I(\mathbf{x}_{k};\tilde{\mathbf{y}}_{k}, \hat{\mathbf{B}}_{kk}).  
\end{equation}
The mutual information is obtained using the differential entropies as
\begin{equation} \label{eq:mutualinformation5}
    I(\mathbf{x}_{k};\tilde{\mathbf{y}}_{k}, \hat{\mathbf{B}}_{kk}) = h(\mathbf{x}_{k}) -h(\mathbf{x}_{k}|\tilde{\mathbf{y}}_{k}, \hat{\mathbf{B}}_{kk}). 
\end{equation}
If we suboptimally choose $\mathbf{x}_{k} \sim \mathcal{N}_{\mathbb{C}}(\mathbf{0},\mathbf{Q}_k)$, then the first term is equal to
\begin{equation} \label{eq:differentialentropyleft5}
    h(\mathbf{x}_{k}) = \log_{2} | \pi e \mathbf{Q}_k |.
\end{equation}
To bound the second term in \eqref{eq:mutualinformation5}, suppose we compute the LMMSE estimate $\hat{\mathbf{x}}_{k}$ of $\mathbf{x}_{k}$ based on $\tilde{\mathbf{y}}_{k}$. We can then upper bound the term using covariance of estimation error as
\begin{align} \nonumber
    h(\mathbf{x}_{k}|\mathbf{y}_{k}, \hat{\mathbf{B}}_{kk}) &=  
    h(\mathbf{x}_{k}-\hat{\mathbf{x}}_{k}|\mathbf{y}_{k}, \hat{\mathbf{B}}_{kk}) 
    \leq h(\mathbf{x}_{k}-\hat{\mathbf{x}}_{k}) \\ &=
    \log_{2} | \pi e  (\mathbf{Q}_{k} - \mathbf{Q}_{k}(\mathbf{Q}_{k}+\mathbf{C}_{\mathbf{n}'_{k}})^{-1}  \mathbf{Q}_{k}  |, \label{eq:differentialentropyright5}
\end{align}
where $\mathbf{C}_{\mathbf{n}'_{k}}$ is the covariance matrix of the term in \eqref{eq:observation_equation_afterZFcombiner}.
By substituting \eqref{eq:differentialentropyleft5} and \eqref{eq:differentialentropyright5} into \eqref{eq:mutualinformation5}, we obtain the expression stated in the theorem, by including the pre-log factor and using the identity $\log_{2}  |\mathbf{I} + \mathbf{A}\mathbf{B}|  = \log_{2}  |\mathbf{I} + \mathbf{B}\mathbf{A}|$. 

\bibliographystyle{IEEEbib}
\bibliography{cellfree}

\end{document}